\documentclass[epsf,graphics,a4,12pt]{article}
\usepackage{epsfig}
\begin{document}
\baselineskip=18pt
\newcommand{\be}{\begin{equation}}
\newcommand{\en}{\end{equation}}
\begin{titlepage}
\begin{center}
{\hbox to \hsize{\hfill .}}
\bigskip
\vspace{6\baselineskip}
{\Large \bf Sampling in AdS/CFT}
\end{center}
\vspace{1ex}
\centerline{\large
L.Alejandro Correa-Borbonet\footnote{correa@fma.if.usp.br,\\
borbonet@ictp.trieste.it}}
\begin{center}
{Instituto de F\'{\i}sica, Universidade de S\~{a}o Paulo,\\
C.P.66.318, CEP 05315-970, S\~{a}o Paulo, Brazil\\
and\\
The Abdus Salam International Centre for Theoretical Physics,\\
Strada Costiera 11, 34100 Trieste, Italy}
%\end{center}
\vspace{4ex}
\large{

{\bf Abstract}\\
}
\end{center}
%\begin{abstract}
\noindent
Recently, it has been proposed by Kempf a generalization of the Shannon 
sampling theory to the physics of curved spacetimes. With the aim of  
exploring the possible links between Holography and Information Theory we argue 
about the similitude of the reconstruction formula in the sampling theo\-ry and the 
bulk-to-boundary relations found in the AdS/CFT context.
%\end{abstract}
\\
\\
\vspace{9ex}
\hspace*{0mm} PACS number(s): 04., 89.70.+c
\vfill
\end{titlepage}

\newpage
%%%%%%%%%%%%%%%%%%%%%%%%%%%%%%%%%%%%%%%%%%%%%%%%%%%%%%%%%%%%%%%%%%%%%%%%%
\section{Introduction.}
%%%%%%%%%%%%%%%%%%%%%%%%%%%%%%%%%%%%%%%%%%%%%%%%%%%%%%%%%%%%%%%%%%%%%%%%%
The search for possible bridges between different areas in science has been 
always a fertile field for researchers. Usually, the implementation of some 
concepts from one area into another sheds light into the features of some 
specific problems. Particularly,  the intersection between Information Theory  
and Physics has inspired many people \cite{demon},\cite{landauer}. 
In this sense Bekenstein \cite{bekenstein} 
pointed out that a century of developments in physics has taught us that 
information is a crucial player in physical systems and processes.  

Within this context Kempf \cite{kempf3} proposed the application of Shannon 
sampling theory \cite{shannon} to generic curved spacetimes. His main idea is that physical 
fields could be constructed everywhere if sampled only at discrete points in 
space. These sampling points should be spaced densely enough, say of the 
order of the Planck distance. Recently he proved \cite{kempf5}
the mathematical conjectures outlined in his previous papers.  One of the 
outputs of this nice idea is that a sampling theoretic ultraviolet 
cutoff at the Planck scale also co\-rresponds to a finite density of 
degrees of 
freedom for physical fields. This allows the holographic principle \cite{susskind} to enter 
in scene. According to t' Hooft and Susskind the combination of Quantum 
Mechanics and Gravity requires the three
dimensional world to be an image of data that can be stored on a two
dimensional projection much like a holographic image. This description
requires only one discrete degree of freedom per Planck area and yet
it is rich enough to describe all three dimensional
phenomena. This bound has not been justified but
it is obvious that the assumption of these ideas
imply a radical decrease in the number of degrees of freedom for
describing the Universe. Maldacena's conjecture on AdS/CFT correspondence 
\cite{maldacena} is the first example realizing such a principle. Subsequently, 
Witten \cite{witten} proposed a precise correspondence between conformal field 
theory observables and those of supergravity on the AdS side.

The purpose of this note is to show, on heuristic grounds, an application of the 
Kempf's idea in the context of AdS/CFT correspondence. We argue about the extreme 
similitude between the bulk-to-boundary formulas in AdS/CFT and the reconstruction 
formula proposed by Kempf. 

%%%%%%%%%%%%%%%%%%%%%%%%%%%%%%%%%%%%%%%%%%%%%%%%%%%%%%%%%%%%%%%%%%%%%%%%%
\section{Scalar amplitudes in AdS/CFT.} \nonumber
%%%%%%%%%%%%%%%%%%%%%%%%%%%%%%%%%%%%%%%%%%%%%%%%%%%%%%%%%%%%%%%%%%%%%%%%%
In this section we briefly review the essential features of bulk-to-boundary 
procedure for a scalar field.

The simplest way is to work in the Euclidean continuation of $\it AdS_{d+1}$
which is the $Y_{-1}>0$ sheet of the hyperboloid \cite{witten},\cite{freedman}:
\be
-(Y_{-1})^{2}+(Y_{0})^{2}+\sum^{d}_{i=1}(Y_{i})^{2}=-\frac{1}{a^{2}}
\en
which has curvature $R=-d(d+1)a^{2}$. The change of coordinates:
\begin{eqnarray}
z_{i}& = & \frac{Y_{i}}{a(Y_{0}+Y_{-1})} ,\\
z_{0}& = & \frac{1}{a^{2}(Y_{0}+Y_{-1})}
\end{eqnarray}
brings the induced metric to the form:
\be
ds^{2}=\frac{1}{a^{2}z^{2}_{0}}(dz^{2}_{0}+d\vec{z}^{2}).
\en
The Euclidean action of the massive scalar field in this background is
\be
S=\frac{1}{2}\int d^{d}zdz_{0} \sqrt{g} \left[ g^{\mu \nu}\partial_{\mu}
\phi \partial_{\nu}\phi+ m^{2} \phi^{2} \right].
\en 
The corresponding wave equation is:
\be
\frac{1}{\sqrt{g}}\partial_{\mu}(\sqrt{g}g^{\mu \nu}\partial_{\nu}\phi)-
m^{2}\phi=0 ,\label{eq:operator}
\en
\be
z^{d+1}_{0}\frac{\partial}{\partial z_{0}}\left [ 
z^{-d+1}_{0}\frac{\partial}{\partial z_{0}}\phi(z_{0},\vec{z})\right]+
z^{2}_{0}\frac{\partial}{\partial \vec{z}^{2}}\phi(z_{0},\vec{z})-
m^{2}\phi(z_{0},\vec{z})=0. \label{eq:motion}
\en
The normalized bulk-to-boundary Green's function
\be
K_{\Delta}(z_{0},\vec{z},\vec{x})=\frac{\Gamma(\Delta)}{\pi^{\frac{d}{2}}
\Gamma(\Delta-\frac{d}{2})}
\left(\frac{z_{0}}{z^{2}+(\vec{z}-\vec{x})^{2}} \right)^{\Delta}
\en
is a solution of (\ref{eq:motion}) with the necessary singular behavior as $z_{0}\rightarrow 0$. Here $\Delta$ is
the largest root of the characteristic  equation of (\ref{eq:motion}).

In the paper \cite{witten} Witten found the solution of (\ref{eq:motion}) that explicitly rea\-lizes the relation between the field 
$\phi(z_{0},\vec{z})$ in the bulk and the boundary configuration $\phi_{0}(\vec{x})$, 
that is given by
\be
\phi(z_{0},\vec{z})=\frac{\Gamma(\Delta)}{\pi^{\frac{d}{2}}\Gamma(\Delta-\frac{d}{2})}\int d^{d}x \left(\frac{z_{0}}{z^{2}+
(\vec{z}-\vec{x})^{2}} 
\right)^{\Delta} \phi_{0}(\vec{x}). \label{eq:bounbulk}
\en

%%%%%%%%%%%%%%%%%%%%%%%%%%%%%%%%%%%%%%%%%%%%%%%%%%%%%%%%%%%%%%%%%%%%%%%%%%%%
\section{Sampling}
%%%%%%%%%%%%%%%%%%%%%%%%%%%%%%%%%%%%%%%%%%%%%%%%%%%%%%%%%%%%%%%%%%%%%%%%%%%%
In an interesting triad of papers \cite{kempf3},\cite{kempf},\cite{kempf2} 
it was pointed out that the mathematical tools 
of the sampling theory could play an important role in the understanding of 
the space-time structure. The sampling theorem states that in order to capture a signal $f(x)$ with bandwidth $\omega_{max}$ for all $x$, it is sufficient to record only the signals' values at the discrete points $\{x_{n}\}$. These sampling points should be spaced densely enough.
In other words, consider the set of square integrable functions $f$ whose frequency content is bounded by $\omega_{max}$. These functions are called band-limited functions. If the amplitudes $\{f(x_{n})\}$ of such a function are known at equidistantly spaced discrete points $\{x_{n}\}$ whose spacing is $\pi/\omega_{max}$ then the function's amplitude $f(x)$ can be reconstructed for all $x$. 
The reconstruction formula is:  
\be
f(x)=\sum^{\infty}_{n=-\infty}  f(x_{n}) \frac{sin[(x-x_{n})\omega_{max}]}{(x-x_{n})w_{max}}
\en
In the article \cite{kempf2} Kempf proposed the generalization of the sampling theory to Riemannian manifolds. There he exposed that the covariant analog of the band-limit is the cutoff of the spectrum of a scalar self-adjoint differential operator. As an explicit example he 
chose the Laplace-Beltrami operator $\Delta$.

Following Kempf we start with the Hilbert space $\cal {H}$ of square integrable scalar functions over the manifold. Later we consider the domain $\cal{D}\subset \cal{H}$, where the Laplacian is essentially self-adjoint. After that we define $P$ as the projector onto the 
subspace spanned by the eigenspaces of the Laplacian with eigenvalues smaller than some fixed maximum value $\lambda_{max}$. If we 
are working with d'Alembertians $\lambda_{max}$ bounds the absolute values of the eigenvalues.  
 
Now let us consider a physical field $|\phi)$. This field belongs to the subspace ${\cal{D}}_{ph}= P.\cal{D}$ where are all the physical 
fields. It is assumed that the field's amplitudes $\phi(x_{n})=(x_{n}|\phi)$ are 
known only at the discrete points $\{x_{n}\}$ of the manifold. While all position 
eigenvectors $|x)$ are need to span $\cal{H}$, sufficiently dense discrete 
subsets $\{x_{n}\}$ of the set of vectors $\{P|x)\}$  can span  ${\cal{D}}_{ph}$. 
The field's coefficients $\{\phi(x_{n})\}$ then fully determine the Hilbert space vector $\vert 
\phi)\in {\cal{D}}_{ph}$ and they determine, therefore, also $(x\vert\phi)$ for all $x$. Namely, defining
$K_{n\lambda}= (x_n\vert\lambda)$, the set of sampling points $\{x_n\}$ is
sufficiently dense for reconstruction iff $K$ is invertible. To see this,
insert the resolution of the identity in terms of the eigenbasis
$\{\vert\lambda )\}$ of $-\Delta$ into $(x\vert\phi)$:
\begin{equation}
(x\vert\phi)=\sum_{\vert\lambda\vert<\lambda_{max}}
\!\!\!\!\!\!\!\!\!\!\!\!\!\!\int ~(x\vert \lambda)(\lambda \vert
\phi)~d\lambda. \label{nueva}
\end{equation}
With K invertible one obtains $(\lambda\vert \phi)=\sum_n K^{-1}_{\lambda,n}
\phi(x_n)$. Substituting back in (\ref{nueva}) we obtain
\be
\phi(x)=\sum_{n}G(x,x_{n})\phi(x_{n})\label{eq:recon}
\en
where
\be
G(x,x_n)~=~\sum_{\vert\lambda\vert<\lambda_{max}}
\!\!\!\!\!\!\!\!\!\!\!\!\!\!\int ~ (x\vert\lambda) K^{-1}_{\lambda,
n}~d\lambda\label{eq:green}
\en
is called the reconstruction kernel.  
%%%%%%%%%%%%%%%%%%%%%%%%%%%%%%%%%%%%%%%%%%%%%%%%%%%%%%%%%%%%%%%%%%%%%%%%%%%%
\section{Boundary to bulk relation as a Sampling mechanism}
%%%%%%%%%%%%%%%%%%%%%%%%%%%%%%%%%%%%%%%%%%%%%%%%%%%%%%%%%%%%%%%%%%%%%%%%%%%%
After reading the previous sections the sharp
reader may have perceived the strong resemblance between the boundary to bulk 
relation (\ref{eq:bounbulk}) and the reconstruction formula (\ref{eq:recon}). 
But we think that this is not a mere formal similarity. Physically speaking the holographic 
principle claims that 
the fundamental degrees of freedom live in 
the boundary. Therefore the degrees of freedom in the bulk could be seen as  
information ``constructed'' from the boundary. The holographic assumption that 
the boundary theory should have only a finite number of degrees of freedom per 
Planck area is also compatible with the sampling condition that requires the 
existence of a dense domain, with spacing of the order of Planck distance, in order 
to do the reconstruction. 

In order to clarify our statement let's consider that the function $\phi$ is 
defined at the discrete points $\vec{x}_{n}$. 
The particularity here lies in the fact that sampling 
points live in the 
boundary. Then the equation 
(\ref{eq:bounbulk}) changes to
\be
\phi(z_{0},\vec{z})=\frac{\Gamma(\Delta)}{\pi^{\frac{d}{2}}\Gamma(\Delta-\frac{d}{2})}
\sum_{\vec{x}_{n}} 
\left(\frac{z_{0}}{z^{2}+(\vec{z}-\vec{x}_{n})^{2}} 
\right)^{\Delta} 
\phi_{0}(\vec{x}_{n}). \label{eq:bounbulksum}
\en
Now, comparing this with equation (\ref{eq:recon}) and treating $\vec{z}$ as a constant we see that they agree if we take  
\be
G(z_{0},\vec{z},\vec{x}_{n})=\frac{\Gamma(\Delta)}{\pi^{\frac{d}{2}}
\Gamma(\Delta-\frac{d}{2})} 
\left(\frac{z_{0}}{z^{2}+(\vec{z}-\vec{x}_{n})^{2}} 
\right)^{\Delta}. \label{eq:ultima}
\en
We have assumed that the differential operator presented in 
(\ref{eq:operator}) has a cutoff in the spectrum. We also suppose that
the reconstruction stability is satisfied.
Therefore the next task will be  to proof (\ref{eq:green}) taking into account (\ref{eq:ultima}).
%%%%%%%%%%%%%%%%%%%%%%%%%%%%%%%%%%%%%%%%%%%%%%%%%%%%%%%%%%%%%%%%%%
\section{Conclusions}
%%%%%%%%%%%%%%%%%%%%%%%%%%%%%%%%%%%%%%%%%%%%%%%%%%%%%%%%%%%%%%%%%%
Potentially Information Theory could play an important role imposing some
constrains to physical theories. Although this relation needs to be 
explored further we showed here a possible implementation of the generalization 
of the Shannon sampling theory to the physics of curved spacetimes, as proposed 
by Kempf. Seeing the sampling procedure as a mechanism that generates the 
bulk's degrees of freedom seems to be a very attractive idea from the holographic 
point of view.  
%%%%%%%%%%%%%%%%%%%%%%%%%%%%%%%%%%%%%%%%%%%%%%%%%%%%%%%%%%%%%%%%%%%%%%%%

 {\bf ACKNOWLEDGMENT:}
The author is grateful to Dr. Achim Kempf for useful comments and Dr. Cesar Castilho for 
useful suggestions and for reading the manuscript.
I would like 
to thank the High Energy Group of the Abdus Salam ICTP for hospitality and support 
during my visit. This 
work was also supported by 
Funda\c{c}\~ao de Amparo \`a Pesquisa do Estado de S\~ao Paulo (FAPESP).

%%%%%%%%%%%%%%%%%%%%%%%%%%%%%%%%%%%%%%%%%%%%%%%%%%%%%%%%%%%%%%%%%%%%%%%%%%%%


\begin{thebibliography}{99}
%%%%%%%%%%%%%%%%%%%%%%%%%%%%%%%%%%%%%%%%%%%%%%%%%%%%%%%%%%%%%%%%%%%%%%%%%%%%
\bibitem{demon} H.S. Leff and A.F. Rex, {\it Maxwell's Demon Entropy, 
Information, Computing} (1990) Princeton University Press.
%%%%%%%%%%%%%%%%%%%%%%%%%%%%%%%%%%%%%%%%%%%%%%%%%%%%%%%%%%%%%%%%%%%%%%%%%%%%%
\bibitem{landauer} R. Landauer, {\it Information is physical}, {\it Physics 
Today}, {\bf 44(5)}(1991) 22.
%%%%%%%%%%%%%%%%%%%%%%%%%%%%%%%%%%%%%%%%%%%%%%%%%%%%%%%%%%
\bibitem{bekenstein} J.D. Bekenstein, {\it Information in the Holographic Universe},
{\it Scientific American} August (2003) 48.
%%%%%%%%%%%%%%%%%%%%%%%%%%%%%%%%%%%%%%%%%%%%%%%%%%%%%%%%%%%%%%%%%%%%%%%%%%%%%
\bibitem{kempf3} A. Kempf, {\it Phys.Rev.Lett.} {\bf 85} (2000) 2873.   
%%%%%%%%%%%%%%%%%%%%%%%%%%%%%%%%%%%%%%%%%%%%%%%%%%%%%%%%%%%%%%%%%%%%%%%%%%%%
\bibitem{shannon} C.E. Shannon, N.J. A. Sloane and A.D. Wyner,
{\it  Claude Elwood Shannon: collected papers} New York, NY., 
The Institute of Electrical and Electronics Engineers, (1993).
%%%%%%%%%%%%%%%%%%%%%%%%%%%%%%%%%%%%%%%%%%%%%%%%%%%%%%%%%%%%%%%%%%%%%%%%%%
\bibitem{kempf} A. Kempf, {\it A Covariant Information-Density Cutoff in 
Curved Space-Time}, gr-qc/0310035.
%%%%%%%%%%%%%%%%%%%%%%%%%%%%%%%%%%%%%%%%%%%%%%%%%%%%%%%%%%%%%%%%%%%%%%%%%%%
\bibitem{kempf2} A. Kempf, {\it Aspects of Information Theory in Curved Space},
qr-qc/0306104.
%%%%%%%%%%%%%%%%%%%%%%%%%%%%%%%%%%%%%%%%%%%%%%%%%%%%%%%%%%%%%%%%%%%%%%%%%%%
\bibitem{kempf5} A. Kempf, {\it On Fields with Finite Information Density}, 
hep-th/0404103. 
%%%%%%%%%%%%%%%%%%%%%%%%%%%%%%%%%%%%%%%%%%%%%%%%%%%%%%%%%%%%%%%%%%%%%%%%%%%
\bibitem{susskind} G. 't Hooft, gr-qc/9310026;
 L.Susskind, {\it J.Math.Phys.} {\bf 36} (1995) 6377.
%%%%%%%%%%%%%%%%%%%%%%%%%%%%%%%%%%%%%%%%%%%%%%%%%%%%%%%%%%%%%%%%%%%%%%%%%
\bibitem{maldacena} J.M. Maldacena, {\it  Adv.Theor.Math.Phys.} {\bf 2} 
(1998) 231.
%%%%%%%%%%%%%%%%%%%%%%%%%%%%%%%%%%%%%%%%%%%%%%%%%%%%%%%%%%%%%%%%%%%%%%%%%%
\bibitem{witten} E. Witten, {\it  Adv.Theor.Math.Phys.} {\bf 2} (1998) 253.
%%%%%%%%%%%%%%%%%%%%%%%%%%%%%%%%%%%%%%%%%%%%%%%%%%%%%%%%%%%%%%%%%%%%%%%%%%%%
\bibitem{freedman} D.Z. Freedman, S.D. Mathur, A. Matusis and L. Rastelli, 
{\it Nucl.Phys.} {\bf B546} (1999) 96.
%%%%%%%%%%%%%%%%%%%%%%%%%%%%%%%%%%%%%%%%%%%%%%%%%%%%%%%%%%%%%%%%%%%%%%%%%%%%
\end{thebibliography}
\end{document}